\documentclass{b98proc}
\include{psfig}
\newcommand{\eqn}[1]{\label{eq:#1}}
\newcommand{\refeq}[1]{(\ref{eq:#1})}
\newcommand{\eq}{eq.~\refeq}

\newcommand{\Eq}{Eq.~\refeq}

\newcommand{\beq}{\begin{eqnarray}}
\newcommand{\eeq}{\end{eqnarray}}
\def\CA{{\cal A}}
\def\Journal#1#2#3#4{{#1} {\bf #2}, #3 (#4)}

\def\NPA{{\em Nucl. Phys.} A}
\def\NPB{{\em Nucl. Phys.} B}

\def\PLB{{\em Phys. Lett.} B}

\def\PRD{{\em Phys. Rev.} D}
\def\PRC{{\em Phys. Rev.} C}

\begin{document}

\title{
EFFECTIVE FIELD THEORY FOR NUCLEAR PHYSICS
}

\author{David B. KAPLAN}
\address{Institute for Nuclear Theory, University of Washington,
  Seattle, WA, 98195-1550, U. S. A.} 


\maketitle

\abstracts{
I summarize the motivation for  the
effective field theory approach to nuclear physics,
and some of its recent accomplishments.  
}

\section{Why effective field theory, why nuclear physics?}

Low energy
data is generally insensitive to the details of interactions at short distance.
It is therefore difficult to learn about
short range interactions; yet by the same token, complete knowledge of the 
physics at short distances is not required for an accurate
understanding of experiments.  Effective field theory exploits this
fact.  The effects of nonlocal interactions at short distance may be
represented  in terms of local operators
in a derivative expansion --- the effective Lagrangian.  The higher an 
operator's dimension, the smaller the effect it has on low energy
physics, and hence one can obtain a useful phenomenological theory by
retaining operators only up to some dimension, fitting their
coefficients to data.  Some effective theories are quite useful, such
as chiral perturbation theory;  some are wildly successful, such as
the standard model of particle physics.  In this talk I will discuss a 
new application currently being developed, nuclear effective theory.

The utility of effective field theory  (EFT) depends on the existence of an
energy gap so that ``short'' and ``long'' distance physics can be
distinguished. It is probably not a useful technique for 
describing turbulence, or protein folding, for example.  In 1990, Weinberg
suggested that nuclear physics could be a subject that would benefit
from an EFT treatment \cite{Weinberg}.  In nucleon-nucleon
interactions, one can identify the low scales to be $m_\pi=140$ MeV,
and the nucleon momentum ($p_F\simeq 280$ MeV in nuclear matter),
while the high scales would be the masses of the vector mesons and
higher resonances.  The $\Delta$ contributes to $NN$ scattering  at a
momentum scale 
$\sim \sqrt{2 M_N (M_\Delta-M_N)}=750$ MeV, comparable to the vector
mesons.  While Weinberg's original proposal has been shown to not be
consistent, a new 
approach has been developed over the past nine months that appears
very promising \cite{KSWa} (see also Ref.~\cite{aleph}). It is this
approach that I will focus in this talk.

What are the benefits of EFT for nuclear physics?
The standard starting point of nuclear theory for the past half century
has been to take the 
$NN$ scattering phase shifts as input, and then  construct a Schr\"odinger
potential that can reproduce those phase shifts. Using this potential
on can solve the Schr\"odinger equation for few body systems, or use
various many-body approximation schemes (Br\"uckner theory, the shell
model, etc.) to solve larger nuclei.  The main weakness of this
approach is that there is no systematic way to improve results, or to
have confidence in their level of accuracy.  The lack of a systematic
expansion becomes evident in many-body calculations when one finds
dependence on ``off-shell amplitudes''.  As is well known, off-shell
amplitudes are arbitrary and can be changed at will by field
redefinitions.  That a purportedly physical quantity depends on the
off-shell definition of an amplitude is a sure sign that a consistent
expansion is lacking. As I will discuss below, a systematic expansion
is obtainable  in the EFT approach to nuclear physics, so that
one never need to specify off-shell matrix elements.

EFT offers other, practical, advantages as well:  by replacing
complicated models of short distance physics by separable contact
interactions (in a justifiable fashion), calculations are vastly
simplified.  In the two nucleon  
sector, many computations have been performed analytically, while in
the three nucleon system, the Fadeev equations are reduced to a one
dimensional integral equation quickly solvable on a desktop
computer.
Furthermore, implementation of relativity as well as chiral and gauge
symmetries is simpler using a local effective Lagrangian, compared to
a more traditional nucleon potential approach.

\section{A systematic expansion}

Developing a systematic expansion for an EFT for nuclear physics is
less obvious than for more familiar applications.  Typically
one introduces all operators allowed by low energy symmetries,
introduces couplings which depend inversely on the high energy scale
$\Lambda$ 
to the power appropriate for the dimension of the operator, and then
compute the Feynman amplitude to the desired power of $p/\Lambda$ or
$m/\Lambda$, where $p$ is a external momentum and $m$ is a light
particle mass.  If performed sensibly,  the computation never
introduces powers of  $\Lambda$ in the numerator and so at which order an
operator enters depends solely upon its dimension.  
 
The
lowest dimension contribution to $NN$ scattering at low energies would 
come from the operator $C_0 (N^\dagger N)^2$ (ignoring spin and
isospin indices) where $N$ is the 
nonrelativistic  nucleon  doublet field and $C_0$ is a coupling constant with
dimension ${\rm (mass)^{-2}}$.
If EFT for nuclear theory worked in the accustomed way, then one would 
expect $C_0\sim 1/(M \Lambda)$ where $\Lambda$ is the vector meson
mass scale, and $M$ is the nucleon mass (see Ref.~\cite{LMa} for a
discussion of the $M$ dependence).  One would then insert this $C_0$
perturbatively:  each insertion would contribute an additional power
of $M p C_0\sim p/\Lambda$ to the amplitude, which is small at low energy.

 The reason why this cannot be the whole story behind $NN$ scattering is 
evident if one looks at the $NN$ phase shifts.  In the ${}^1S_0$
channel $np$ scattering exhibits a scattering length $a
= -1/(8\, {\rm MeV})$, while in the ${}^3S_1$ channel the $np$
scattering length  is $a=1/(40\, {\rm MeV})$.  These length scales are 
obviously neither the inverse pion or vector meson masses, but must
arise through nonperturbative dynamics.  In the potential model
approach, they are due to a near cancellation between potential and
kinetic energy. The effective field theory description instead resembles a 
condensed matter system near a phase transition, a system that
similarly exhibits length scales (correlation lengths) much longer
than the fundamental scale of the system.  In other words, one will
have a quantum field theory tuned to lie near a nontrivial fixed
point, where the fixed point corresponds to infinite scattering
length. The EFT is then necessarily nonperturbative, and the actual dimension 
of operators can be different than their naive dimension, with power
law corrections to the scaling of coupling constants.

To exhibit this fixed point behavior, consider $NN$ scattering at
momenta far below $m_\pi$.  The scattering amplitude is then well described 
solely in terms of the scattering length $a$,
\beq
\CA \simeq  -{4\pi\over M}{1\over 1/a + ip}\ ,\qquad p\equiv
\sqrt{ME_{cm}}\ .
\eqn{amp}
\eeq
This can be reproduced by the effective theory consisting of
nonrelativistic nucleons interacting via the $C_0(N^\dagger N)^2$
term.  By iterating bubble diagrams with a $C_0$ 
at each vertex, one finds
\beq
\CA_{eff} = {-1\over {1\over C_0}-  I(p)}\ ,
\eeq
where $I(p)$ is the linearly divergent loop integral
\beq
I(p) = \int {d^3q\over (2\pi)^3} {M\over p^2-q^2}\ .
\eeq
If one evaluates this diagram using dimensional regularization and the 
$PDS$ subtraction scheme introduced in Ref.~\cite{KSWa} (or similarly, if
one performs a momentum space subtraction at $p^2=-\mu^2$
,\cite{Weinberg,Geg1,MS}) one finds 
that in terms of the renormalized coupling $C_0(\mu)$, the amplitude
is
\beq
\CA_{eff} =- {4\pi\over M} {1\over {4\pi\over M   C_0(\mu)} + \mu+ip }\ 
\eeq
Comparison with \eq{amp}
 implies 
\beq
C_0(\mu) = {4\pi\over M}{1\over -\mu + 1/a}\ .
\eeq
Note that the dimensionless coupling constant defined as $\hat C_0
=- M\mu C_0/(4\pi)$, satisfies the
renormalization group equation
\beq
\mu {d \hat C_0\over d\mu} = \hat C_0 (1-\hat C_0)\ .
\eeq
This equation features a nontrivial  UV fixed point at $\hat C_0=1$,
corresponding to $1/a=0$.  At this fixed  point $C_0(\mu)\propto
1/\mu$, so that in fact   $C_0(N^\dagger N)^2$ behaves as a marginal
operator, instead of irrelevant.

We can now do some reverse engineering and identify a consistent power 
counting scheme: Since $C_0(\mu)\propto 1/\mu$, and each loop brought
with it a factor of $\mu$ or $p$, we can define an expansion variable
$Q$ with $\mu \sim p\sim Q$.  Then we see that every term in the bubble 
sum we just performed was order $Q^{-1}$.  This is our leading order
amplitude. The final expression \Eq{amp} implies that $1/a\sim Q$;
our expansion is only valid for systems with large scattering
length. 

To proceed to higher order we consider higher derivative contact
interactions
\beq
C_0 + C_2 p^2 + C_4 p^4 +\ldots
\eeq
as well as interactions involving pions and
photons.  Simple power counting rules emerge, allowing one to
systematically expand the amplitude (the KSW expansion): 
\footnote{Rule (1.) applies to $s$-wave scattering; for a more
  detailed discussion of power counting including arbitrary partial waves,
  see Ref.~\cite{KSWb}.  The second option in rule (4.) occurs at NNLO
  where radiation pions first appear.}
\begin{enumerate}
\item $C_{2n} p^{2n} \sim Q^{n-1}$;
\item $m_\pi\sim Q$;
\item $\nabla \sim Q$, $\partial_t\sim Q^2$ for derivative couplings
  of the pion or photon; 
\item Loop integrations $d^4q$  count as $\sim Q^5$ if $q_0\sim
  q^2/M$, otherwise as $Q^4$;
\item propagators count as $Q^{-2}$.
\end{enumerate}
Armed with these rules, one computes physical observables to a given
order in $Q$.  By working consistently to a given
order, one is ensured of always being able to renormalize the theory
(one has enough couplings to absorb divergences) and no observable
will depend on off-shell matrix elements. 

The above rules show that effective nuclear theory contains chiral
perturbation theory with its usual perturbative treatment of the
pion...one never needs to 
sum up pion exchange to all orders.  That is useful in practice, as it
allows for analytical calculations.  It is also satisfying
theoretically, since it would be hard to imagine chiral perturbation
theory to be relevant if one needed to treat pions nonperturbatively.

\section{Application to a toy model}
\label{sec:3}

Before proceeding to the real world, it is worthwhile to analyze a
toy model that exhibits two distinct length scales
and which can be solved analytically.
That toy model is a pair 
of nucleons interacting via delta shell potentials
\beq
V(r) = g_\rho {m_\rho\over M}\delta(r-1/m_\rho) + g_\pi {m_\pi\over
  M}\delta(r-1/m_\pi)\ .
\eqn{twods}
\eeq
The couplings $g_\rho$, $g_\pi$ are dimensionless and normalized so
that $g=-1$ is a strong coupling just capable of producing a bound
state. In order to have a power counting similar to what is found in
the real world, I define $g_\pi\equiv - m_\pi/\Lambda$, where $\Lambda$
has dimensions of mass.  

The phase shift $\delta$ may be computed
analytically in this 
model, and can be taken to be a function of the five variables
$\left\{m_\rho, \Lambda, m_\pi, a, p\right\}$, where $a$ is the
scattering length and $p=\sqrt{ME_{cm}}$. If we perform the rescaling
\beq
m_\pi\to \epsilon m_\pi\ ,\qquad p\to \epsilon p\ \qquad a\to {1\over
  \epsilon} a\ ,
\eeq
and expand the phase shift in powers of $\epsilon$ as
\beq
\delta(m_\rho,\Lambda,\epsilon m_\pi, a/\epsilon,\epsilon p) =
\delta_0 + \delta_1 \epsilon + {1\over 2} \delta_2\epsilon^2 +\ldots
\eqn{exp}
\eeq
then, following the discussion in the previous section, we would
expect  the functions $\delta_n(m_\rho,\Lambda,m_\pi,a,p)$ to
correspond to the $n^{th}$ term in the KSW expansion of the effective
theory.
In fact, this is correct.  When the short range part of the
interaction in \Eq{twods} is replaced by contact interactions,
\beq
V_{eff} = C_0+ C_2 p^2 + C_4 p^4 + \ldots +  g_\pi {m_\pi\over
  M}\delta(r-1/m_\pi)\ , 
\eeq
and one expands the amplitude using the rules of the previous section, 
one finds that appropriate choices of the $C_{2n}$ couplings allow one 
to recover the expansion \Eq{exp} of the exact result, order by order
\footnote{The procedure involves an expansion of $C_0$ as discussed in 
  Refs.~\cite{MS,CH}.}.
\begin{figure}[ht]
\vbox{\centerline{\psfig{figure=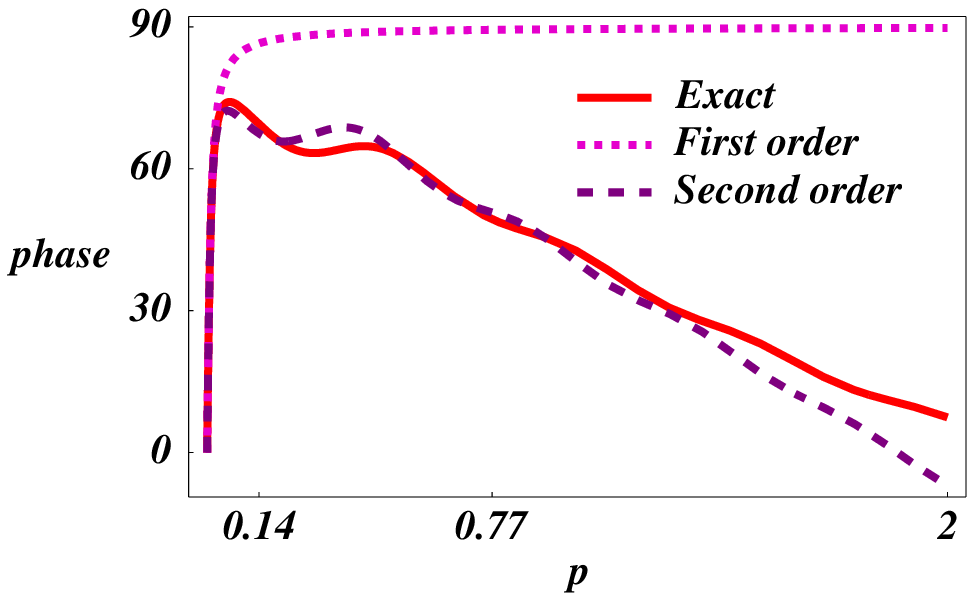,width=9cm}}
\caption{Leading (dotted)and subleading (dashed) EFT result for the
  toy model, compared with the exact answer (solid),  plotted
  as degrees vs. $p=\sqrt{ME_{cm}}$ in GeV. }
\label{comp12}
\centerline{\psfig{figure=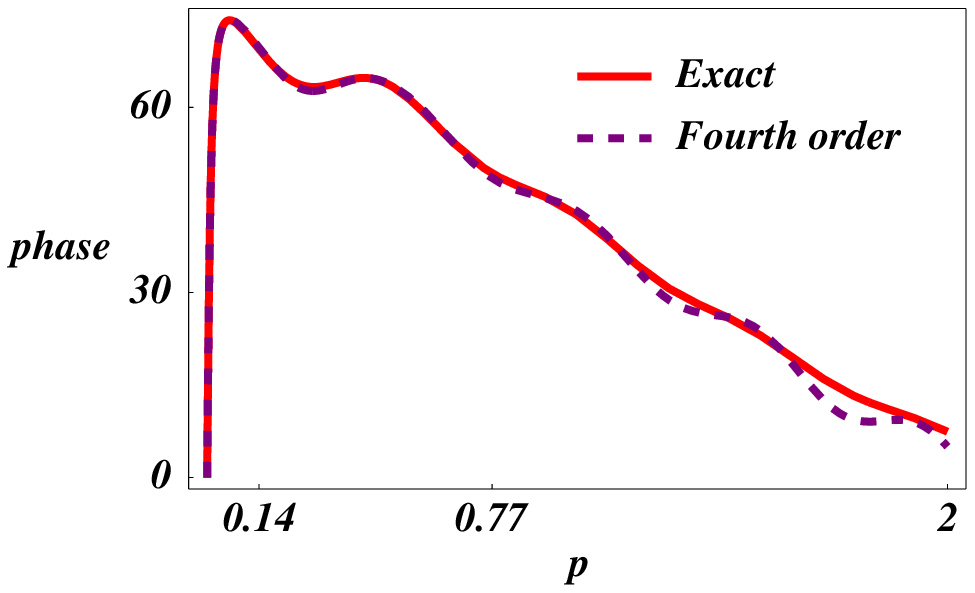,width=9cm}}
\caption{Fourth order EFT result (dashed) compared with the exact
  result (solid), plotted
  as degrees vs. $p=\sqrt{ME_{cm}}$ in GeV. }
}
\label{comp4}
\end{figure}

How quickly does the series converge to the
true phase shift?
I take $m_\pi=140$ MeV, $m_\rho=770$ MeV.  In order to have the toy
model resemble realistic $np$ scattering in the ${}^1S_0$ channel, I
choose $a=-23$ fm, and set  $\Lambda=500$ MeV so that the ``pion''
interaction in the toy model resembles the true one-pion exchange
potential \footnote{By choosing $\Lambda=500$ MeV, we get
  $g_\pi=-m_\pi/\Lambda=-0.28$, which is $28\%$ of the critical value
  that would give rise to a boundstate due to pion exchange alone; the 
  real one-pion exchange Yukawa potential in the ${}^1S_0$ channel  has a
  strength which is similarly 28\% of the critical value.
  Furthermore, for both the toy and realistic pion potentials, the maximum
  phase shift in the absence of additional interactions is  $\sim
  12^\circ$.}.   
In Fig.~\ref{comp12} I have plotted the first two terms in the KSW
expansion (dashed lines) against the exact result (solid line).  The
lowest order (LO) result gets the scattering length exactly, but quickly
fails to describe the ``data''. The next-to-leading order (NLO) result 
is a vast improvement, working crudely up to $p\sim m_\rho$. In
Fig.~\ref{comp4} I show the  result at fourth order in the KSW
expansion; it exhibits excellent agreement up to $p\sim m_\rho$, and
shows not only that the expansion is converging rapidly (in spite of
the enormous difference between LO and NLO results), but also that the 
radius of convergence in momentum is set by $m_\rho$ and not $m_\pi$.
\begin{figure}[t]
\centerline{\psfig{figure=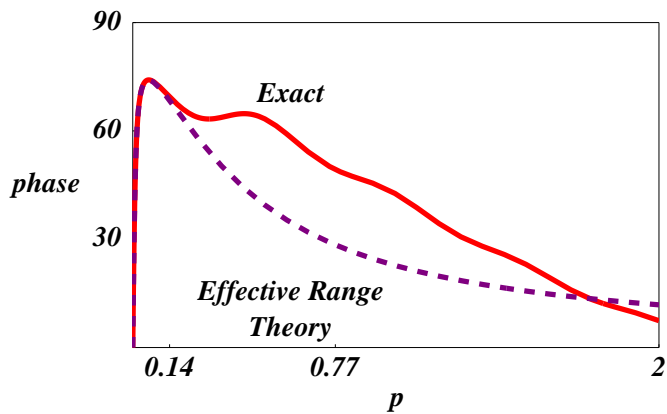,width=9cm}}
\caption{The effective range approximation for the phase shift in the
  toy model (dotted) compared with the exact result (solid), plotted
  as degrees vs. $p=\sqrt{ME_{cm}}$ in GeV.  The approximation breaks
  down at $p\sim m_\pi$. } 
\label{effr}
\end{figure}
For comparison, I have plotted the effective range result in Fig.~\ref{effr}.
It does an excellent job at describing the true phase shift for $p<
m_\pi$ (better than the NLO EFT result), but fails as one would expect 
above $p\sim m_\pi$  

Recently several papers have appeared
suggesting that pions 
must be included nonperturbatively in order to obtain convergent
answers above $p\sim m_\pi$ \cite{Geg2,CH,FS}.  This does not appear to
be the case in 
the toy model, and I don't believe it is true for the real world.  Instead, I
think one gets into trouble by trying to exactly reproduce the effective 
range expansion parameters $a$ and $r_0$ at finite order in the KSW
expansion. That is because one is forcing the short range physics to
account for the contributions to these observables from neglected
higher order pion exchange.  By incorrectly determining the short
distance physics (which is finely tuned), one runs into trouble at
higher momentum.   In the toy model considered here, choosing the $C_{2n}$
coupling to exactly reproduce the $(n+1)$ term in the effective range
expansion yields an EFT prediction that fails (at any order in the
expansion) at $p\sim m_\pi$.   These
issues are to be discussed at greater length in Ref.~\cite{KS}.

\section{The real world}
\label{sec:4}

\begin{figure}[t]
\centerline{\psfig{figure=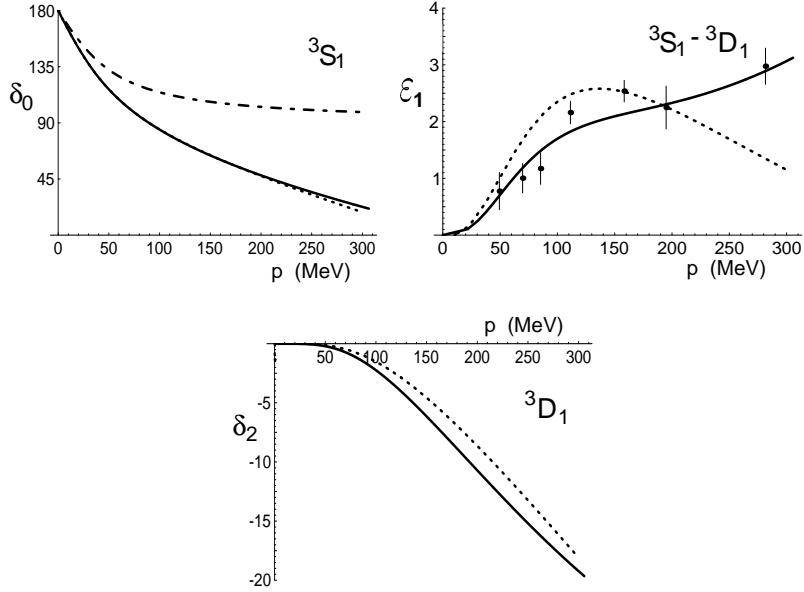,width=11cm}}
\caption{$NN$ phase shifts from Ref.~\protect\cite{KSWa}. The dot-dash and
  dotted lines are the 
 LO and NLO  EFT results respectively.  The solid lines are from the
 Nijmegen partial wave (multi-energy) analysis;  the data points for
 $\epsilon_1$ are from a single-energy analysis \protect\cite{Nij}.  }
\label{phases}
\end{figure}

\begin{figure}[ht]
\centerline{\psfig{figure=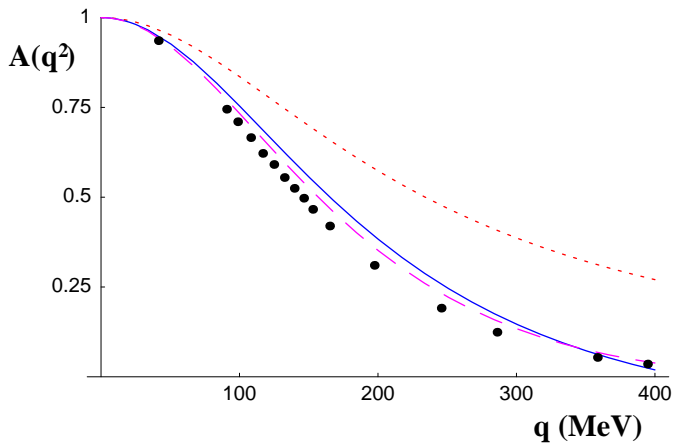,height=6cm}}
\caption{Electric form factor of the deuteron.  The dotted and solid
  curves are the LO and NLO effective field theory results
  respectively;  the dashed curve is the effective range theory
  prediction. The data are
  from various sources \protect\cite{KSWb}. }
\label{emda}
\end{figure}

\begin{figure}[ht]
\centerline{\psfig{figure=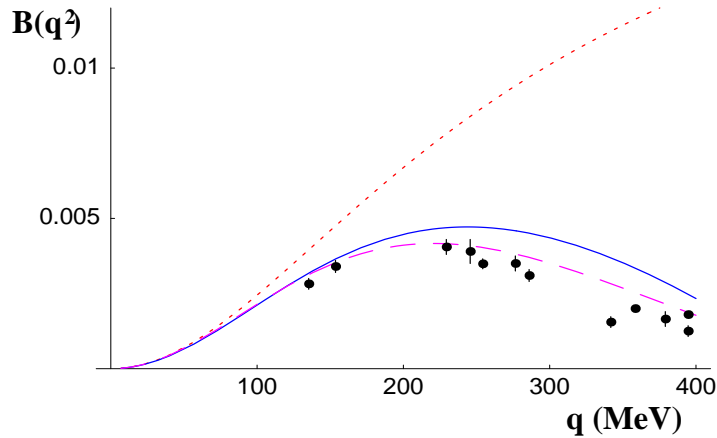,height=6cm}}
\caption{Magnetic form factor of the deuteron.  The dotted and solid
  curves are the LO and NLO effective field theory results
  respectively;  the dashed curve is the effective range theory
  prediction \protect\cite{KSWb}.}
\label{emdb}
\end{figure}

I will now briefly describe the status of calculations for real two and
three nucleon physics using the KSW expansion. Phase shifts have been
fit to NLO \cite{KSWa} in both the ${}^1S_0$ and ${}^3S_1-{}^3D_1$
channels, as shown in Fig.~\ref{phases}.
Using the $C$ parameters derived from the $NN$  phase shifts, a number of
two-nucleon processes have been computed.  In
Figs.~(\ref{emda},\ref{emdb}) I show results 
for the electromagnetic form factors of the deuteron from
Ref.~\cite{KSWb}.  
\begin{figure}[ht]
\centerline{\psfig{figure=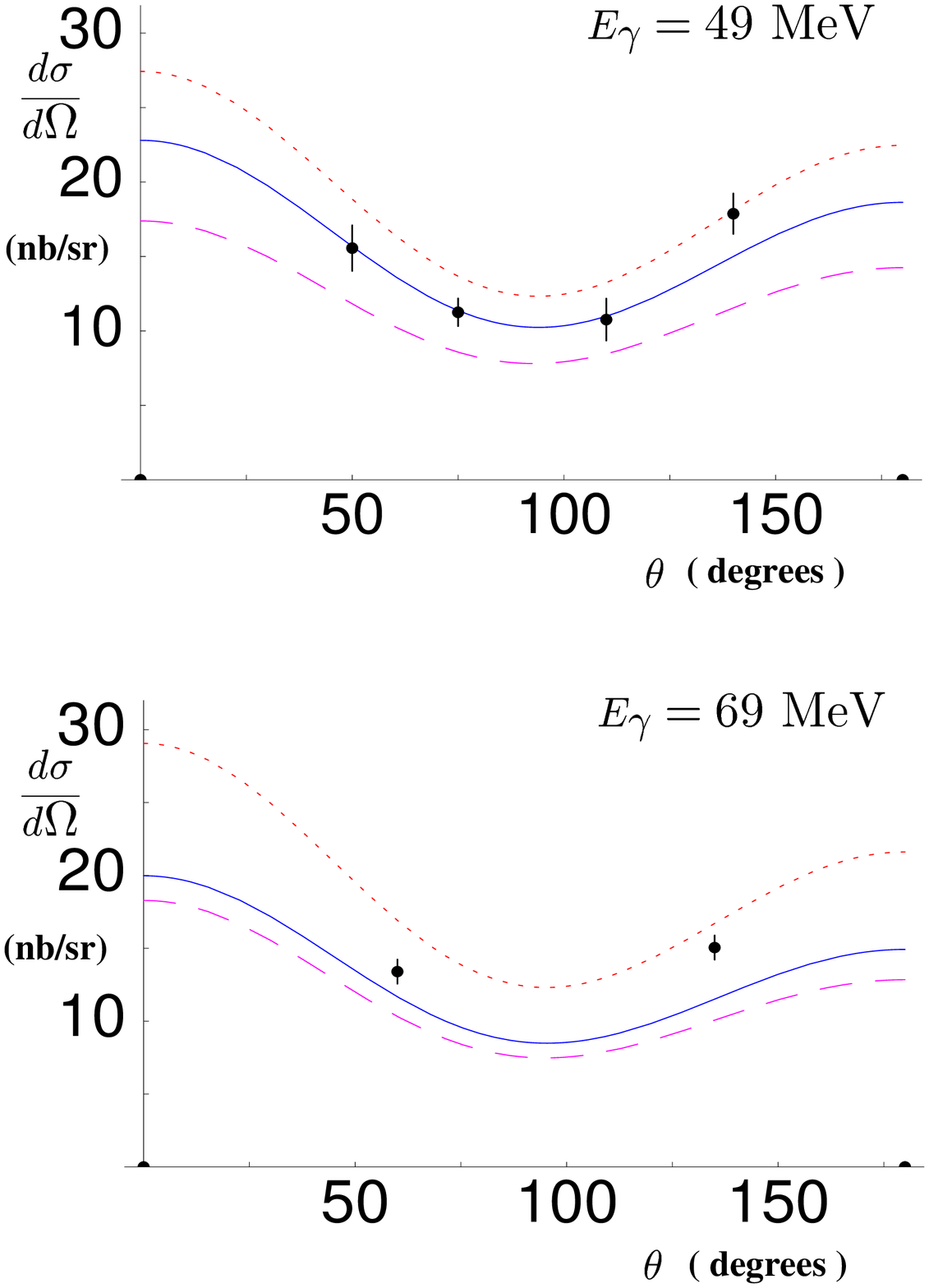,height=12cm}}
\caption{Angular distribution for Compton scattering off the deuteron, 
  for two different energies. In each case the dashed (red) curves represent 
  the LO result; the solid (blue) curves are the full NLO result.
  The dotted (red) curves are the NLO result when nucleon
  polarizability graphs are omitted.  From Ref.~\protect\cite{compton}. }
\label{compt}
\end{figure}
These figures give evidence that the KSW expansion is converging.  However, it
also shows that conventional effective range theory gives a somewhat
better fit to the data than does the EFT
calculation at NLO.  Presumably this is because the effective range
approximation is fit to reproduce low energy scattering data exactly,
while  the EFT calculation yields low energy amplitudes in an
expansion in powers of 
$\sim m_\pi/\Lambda\sim 1/3$. 

\begin{figure}[t]
\vbox{
\centerline{\psfig{figure=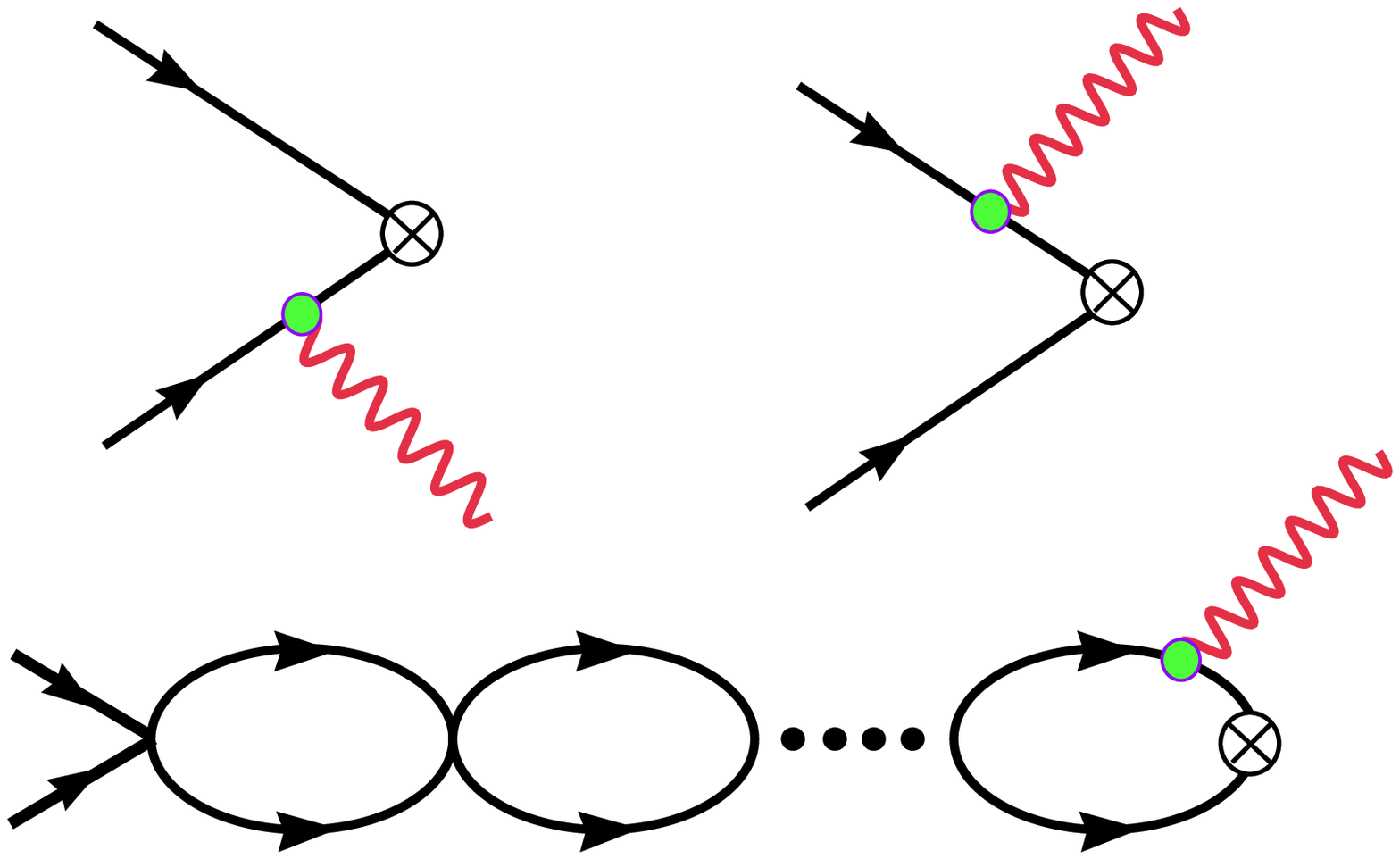,width=7cm}}
\centerline{\psfig{figure=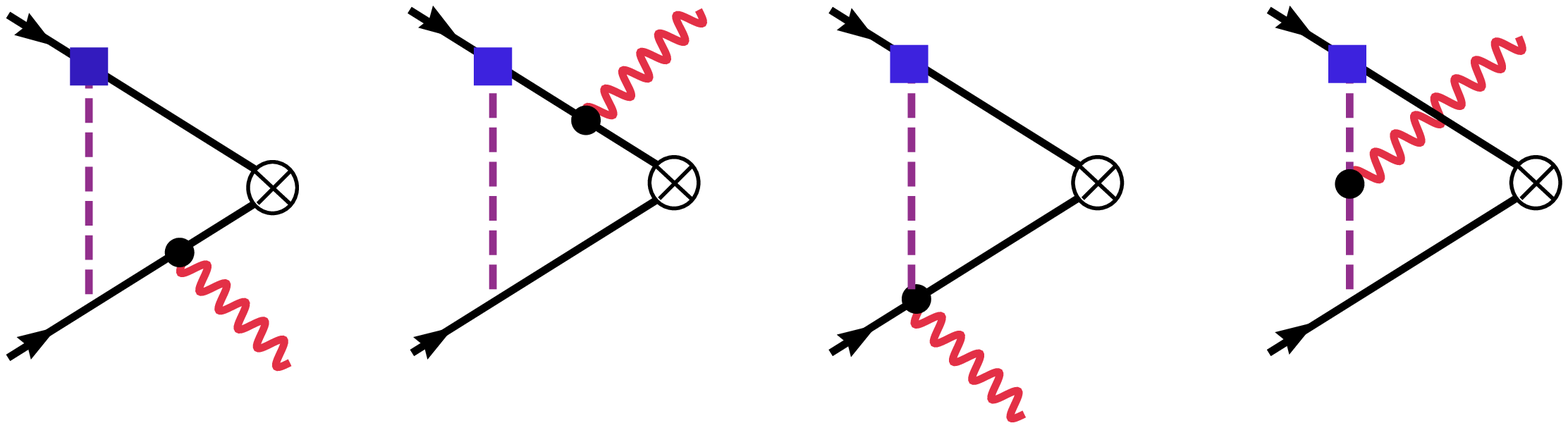,width=7cm}}
\centerline{\psfig{figure=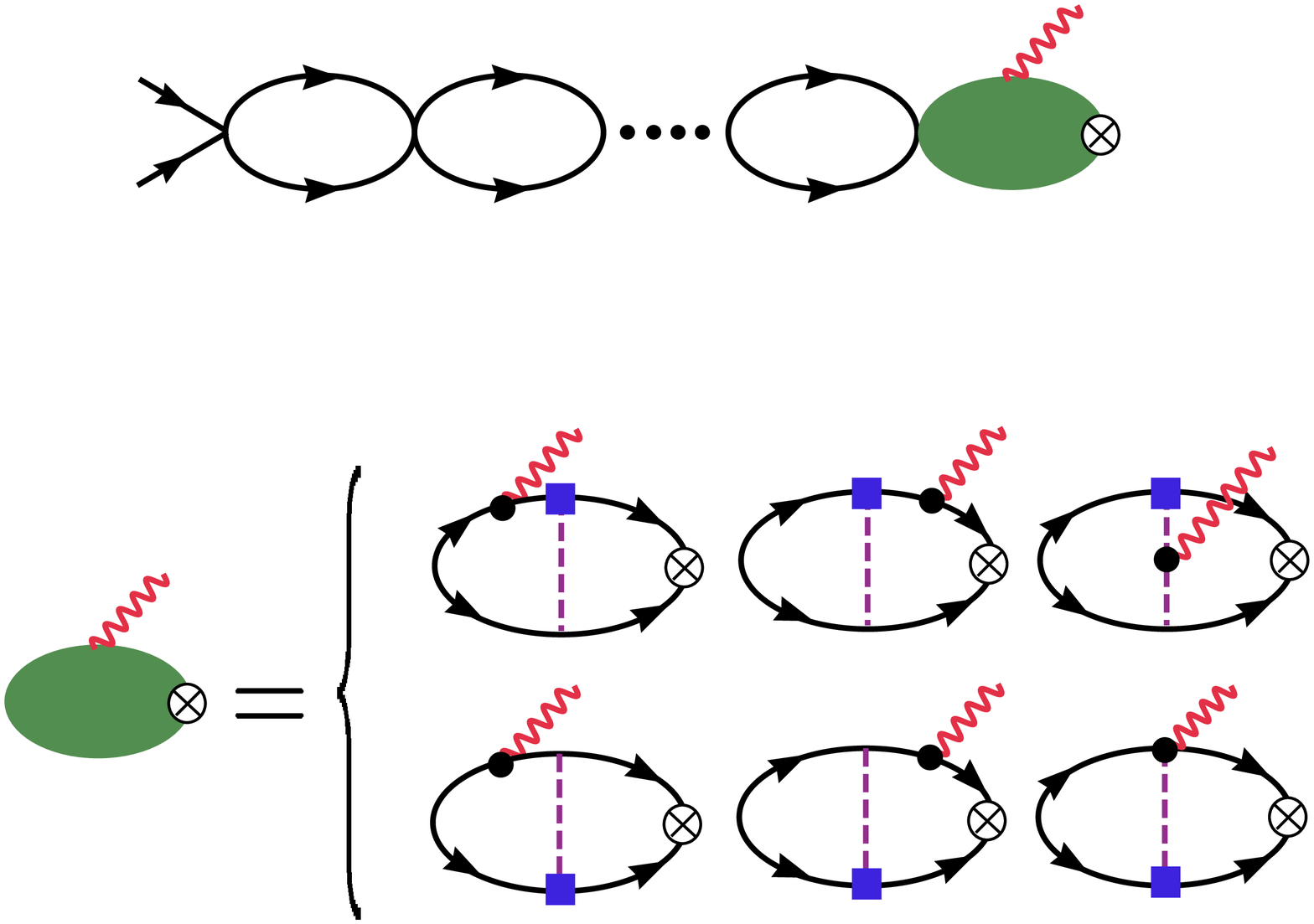,width=7cm}}
\caption{ Feynman diagrams contributing to parity violating $np$
  capture to lowest order,  from Ref.~\protect\cite{KSSW}.  The square vertex
  arises from the weak interactions.}}
\label{npgraphs}
\end{figure}

In Fig.~\ref{compt} is displayed the result from Ref.~\cite{compton} for
Compton scattering off the deuteron.
Recently $np$ capture has been examined, both the parity conserving
part to NLO \cite{SSW}, and the parity violating part to LO
Ref.~\cite{KSSW}. All of these calculations have been performed
analytically.  To give you an idea what is involved, I display in
Fig.~\ref{npgraphs}  the
graphs that had to be computed for the calculation of parity violation 
in $np$ capture.
Other two-nucleon EFT calculations using the KSW expansion include the 
deuteron anapole moment\cite{anapole} and polarizability\cite{polar}.

\begin{figure}[t]
\centerline{\psfig{figure=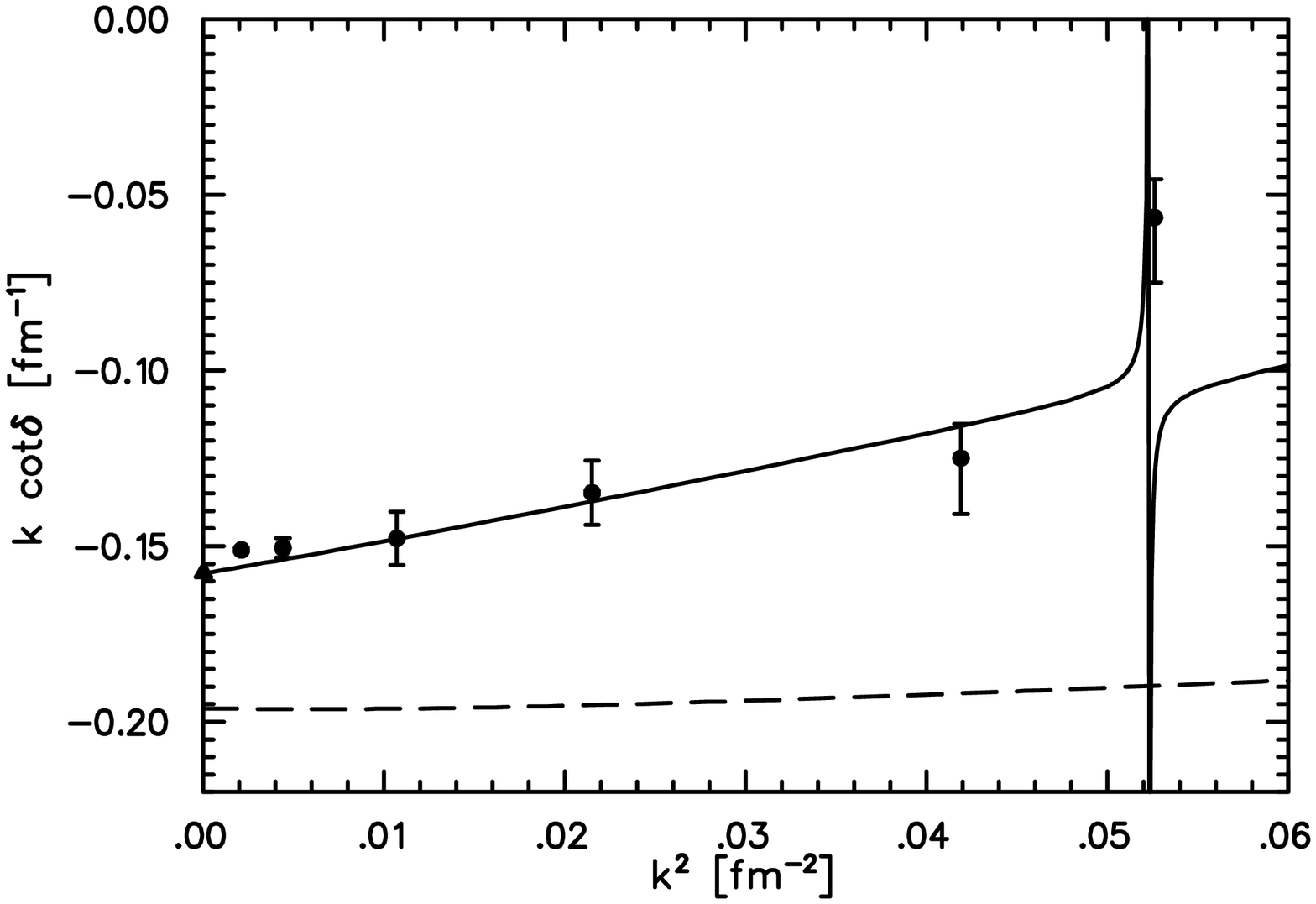,angle=90, width=12cm}}
\caption{$N-d$ scattering phase shift, $J=3/2$ channel, computed in
  EFT.  The only input are the two-body $J=1$ scattering length and
  effective range. From Ref.~\protect\cite{BVKb}. }
\label{ndscat}
\end{figure}

The most interesting challenge  in nuclear effective theory presently is
to extend the analysis to nuclear matter or finite nuclei.  There are
a number of issues that arise in the three-body system that must
 be understood before progress can be made.  The simplest
three-nucleon process to consider is $Nd$ scattering in the $J=3/2$
channel. This was recently computed in Refs.~\cite{BVKa,BVKb} and the
result is displayed in Fig.~\ref{ndscat}.  Great progress has also been made in
understanding the more interesting $J=1/2$ channel from the EFT
perspective\cite{BVKc}. The 
problem is that the graphs one must compute must be represented as an
integral equation, and no systematic power counting has yet been developed , although many interesting features have been understood.
Understanding this system promises to be rewarding.

\section{Discussion}
\label{sec:5}

How does the EFT approach measure up to conventional nuclear physics,
where one solves the Schr\"odinger equation with a phenomenologically
derived potential?  So far, the LO and NLO calculations performed in the past
nine months are not of comparable accuracy.
Potential models obviously do much better than EFT at fitting $NN$
phase shifts, as  
they typically have about two dozen parameters expressly tuned to fit
the data. What about other quantities?
\begin{itemize}
\item{\it Deuteron electric moments:}
 State-of-the-art potential models also do quite well at explaining quantities
particularly sensitive to long distance physics.  This includes the deuteron
charge radius and magnetic moments, accurate to $\sim 1\%$ \cite{PPCPW}.  In
contrast, the EFT predictions for the deuteron charge radius are off
by $28\%$ at LO, and by $11\%$ at NLO \cite{KSWb}.  The reason for this rather
poor agreement is that even low momentum properties get corrections
order by order in the KSW expansion. The deuteron quadrupole moment is 
off by $6\%$ in potential model
calculations\cite{PPCPW}; only the LO result from EFT has been
obtained to date, and it is off by $39\%$. 
\item{\it Deuteron polarizability:} The
polarizability of the deuteron has not been measured, but different
potential models agree with each other on the value of the scalar
electric polarizability to $\sim 0.3\%$ \cite{FP};  the EFT
calculation differs from the potential model  results by $\sim 39\%$
at LO, and $6\%$ at NLO. 
\item{\it Deuteron Compton scattering:}  Compton scattering in the
  energy range shown in Fig.~\ref{compt} is less sensitive to the very 
  long range features of the deuteron wavelength. Predictions from
  potential models \cite{potcom} agree with the data at the
$\sim 10\%$ level, which is comparable to the agreement found in the
NLO EFT calculation.
\end{itemize}
My conclusion is that the operative expansion parameter in the KSW
expansion is $\sim 1/3$, with LO results agreeing with data to the
30\% level and NLO agreeing to the 10\% level.  NLO calculations are
typically not very arduous, and can usually be obtained in closed,
analytic form. An important NLO calculation yet to be performed is the 
quadrupole moment of the deuteron. I expect that NNLO calculations,
expected to be accurate to $\sim 3\%$,
will prove to be 
competitive with --- and in some cases, superior to --- potential
model calculations for a wide variety of 
two nucleon processes.  Such calculations involve relativistic
corrections and propagating pions, and have not yet been fully
understood, although several groups are working on the problem.  NNLO
results most competitive with potential model results will be those
relatively insensitive to the long wavelength tail of the deuteron
wave function. 
Perhaps a better understanding of how in practice to match the KSW
expansion to the effective range expansion at very low energies could
improve the predictions for the latter observables.  

Extending the EFT approach to few- and many-body
systems remains an intriguing challenge, one that must be surmounted
before EFT can really be of much use in nuclear theory.
Like civilizations, physics theories progress from barbarism, to
civilization to decadence (occasionally skipping the civilized
state).  Nuclear effective theory is still in the barbaric stage.
Nevertheless, great progress has been made over the past nine months,
and I am optimistic that the techniques will catch on and eventually
prove themselves invaluable\footnote{Obviously this has been a very
  personal perspective on the subject; 
for other views on how the effective field theory technique should be
applied to nuclear physics, the reader
should  consider the abundant references found in Ref.~\cite{CH}, as
well as the recent review by Rho \cite{Rho}.}.


\end{document}